\documentclass[12pt]{article}
\usepackage[totalwidth=480pt, totalheight=680pt]{geometry}
\usepackage{amsmath}
\usepackage{amssymb}
\usepackage{graphicx} 
\usepackage{url} 
\usepackage{subfigure}
\usepackage{color}

\newcommand{\bbe}{\begin{empheq}[box=\fbox]{equation}}
    \newcommand{\bbea}{\begin{empheq}[box=\fbox]{align}}

        \newcommand{\rT}{r_{s}} \newcommand{\rR}{r_{g}}
        \newcommand{\cod}{_{\text{co}}} \newcommand{\cad}{_{\text{ca}}}

        \newcommand{\Tx}{\ensuremath{_\text{Tx}}}
        \newcommand{\Rx}{\ensuremath{_\text{Rx}}}

        \newcommand{\LO}{\ensuremath{_\text{L.O.}}}
        \newcommand{\beat}{\ensuremath{_\text{b}}}
        \newcommand{\beatc}{\ensuremath{_{\text{b}_0}}}

        \newcommand{\co}{\ensuremath{^\text{co}}}
        \newcommand{\ca}{\ensuremath{^\text{ca}}}

         \newcommand{\vxR}{\vec{x}_{g}}
        \newcommand{\vxT}{\vec{x}_{s}} \newcommand{\Dtc}{\Delta \tau_\text{mo}}
         \newcommand{\Ol}{\mathcal{O}}

\title{Atomic Clock Ensemble in Space (ACES) data analysis}

\author{F~Meynadier$^1$, P~Delva$^1$, C~le~Poncin-Lafitte$^1$, C Guerlin$^{2,1}$,
P~Wolf$^1$} 

\date{}

\begin{document}

\maketitle

{$^1$LNE-SYRTE, Observatoire de Paris, PSL Research
University, CNRS, Sorbonne Universit\'es, UPMC Univ. Paris 06, 61 avenue de
l'Observatoire, 75014 Paris, France} 

{$^2$Laboratoire Kastler Brossel,
ENS-PSL Research University, CNRS, UPMC-Sorbonne Universit\'es, Coll\`ege de
France, 75005 Paris, France} 

Contact : \texttt{Frederic.Meynadier@obspm.fr}

\vspace{10pt} September 2017, revised version December 2017 

Accepted in Classical and Quantum Gravity, 2018, Volume 35, Number 3 :

\url{https://doi.org/10.1088/1361-6382/aaa279}

\begin{abstract} The Atomic Clocks Ensemble in Space (ACES/PHARAO mission, ESA
    \& CNES) will be installed on board the International Space Station (ISS)
    next year. A crucial part of this experiment is its two-way MicroWave Link
    (MWL), which will compare the timescale generated on board with those
    provided by several ground stations disseminated on the Earth. A dedicated
    Data Analysis Center (DAC) is being implemented at SYRTE -- Observatoire de
    Paris, where our team currently develops theoretical modelling, numerical
    simulations and the data analysis software itself.

In this paper, we present some key aspects of the MWL measurement method and
the associated algorithms for simulations and data analysis. We show the
results of tests using simulated data with fully realistic effects such as
fundamental measurement noise, Doppler, atmospheric delays, or cycle
ambiguities. We demonstrate satisfactory performance of the software with
respect to the specifications of the ACES mission. The main scientific product
of our analysis is the clock desynchronisation between ground and space clocks,
i.e. the difference of proper times between the space clocks and ground clocks at participating institutes. While in flight, this
measurement will allow for tests of General Relativity and Lorentz invariance at unprecedented levels, e.g. measurement of the gravitational redshift at the $3\times10^{-6}$ level.

As a specific example, we use real ISS orbit data with estimated errors at the
10 m level to study the effect of such errors on the clock desynchronisation
obtained from MWL data. We demonstrate that the resulting effects are totally
negligible.  \end{abstract}

%
\vspace{2pc} \noindent{\it Keywords}: General Relativity, Atomic Clocks, Tests
of Fundamental Physics, Relativistic Time and Frequency Transfer, Data
Analysis, Space Mission\\

\section{Introduction}

The Einstein Equivalence Principle (EEP) is the foundation of General
Relativity (GR) and more generally of all metric theories of gravitation
\cite{Will2006}, and therefore requires extensive experimental verification.
Additionally, numerous theories beyond GR that seek to unify gravitation with
other fundamental interactions in nature allow for a violation of EEP at some,
a priori unknown, level. Although the different aspects of the EEP like the
Universality of Free Fall (UFF) or the gravitational redshift of clocks are
related \cite{Will2006}, that relationship cannot be quantified a priori
\cite{Wolf2016} and therefore they need to be tested independently. UFF is
currently being tested at unprecedented uncertainty in the MICROSCOPE space
experiment \cite{Touboul2012} that was launched in 2016. Another famous aspect
of the EEP is the gravitational redshift of clocks, that is so far best tested
in the Gravity Probe A experiment \cite{Vessot1979,Vessot1980,Vessot1989} that
used a hydrogen maser onboard a rocket on a parabolic trajectory which was
compared via a two-way microwave link to a hydrogen maser on the ground. One of
the main scientific objectives of the ACES/PHARAO space mission is to improve
on that test by roughly two orders of magnitude.

The ACES/PHARAO mission is an international metrological space mission led by
the European Space Agency (ESA) in collaboration with the French Space Agency
(Centre National d'\'Etudes Spatiales, CNES). It aims at realizing a time scale
of high stability and accuracy on board the International Space Station~(ISS)
using a unique cold-atom space clock developed by CNES in collaboration with
LNE-SYRTE, together with a space hydrogen maser (SHM). Additionally, ACES will
be equipped with microwave (MWL) and optical (European Laser Timing, ELT) links
which will allow frequency/time comparisons between the onboard time scale and
time scales on the ground.

The on-board timescale will track the Space Station clock's proper time, accumulating
phase difference with respect to ground-based counterparts as it navigates
through the Earth’s gravity potential at orbital speed. Keeping track of proper
time differences, i.e. clock desynchronisation, allows to test General
Relativity predictions : such tests are then limited by the noise introduced by
the clocks and the link between them. In the case of ACES/PHARAO, the MWL has
been designed to suit the performance of the space clock during ISS visibility
passes. 

\begin{figure}[h] \begin{minipage}[h]{.45\linewidth}
        \includegraphics[width=\linewidth]{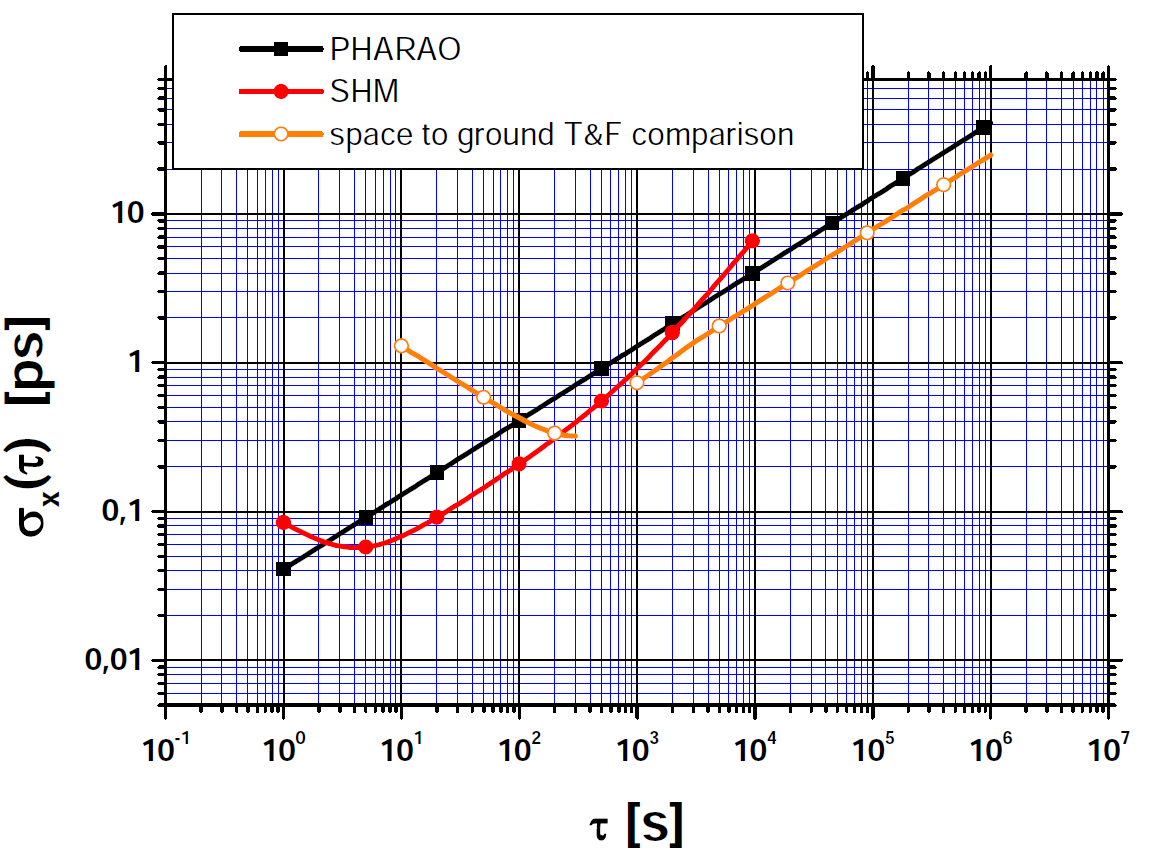}
        \caption{\label{f:sigma_x}Performance objective of the ACES clocks and
    the ACES space-ground time and frequency transfer expressed in time
deviation.}
        
    \end{minipage} \begin{minipage}[h]{.45\linewidth}
        \includegraphics[width=\linewidth]{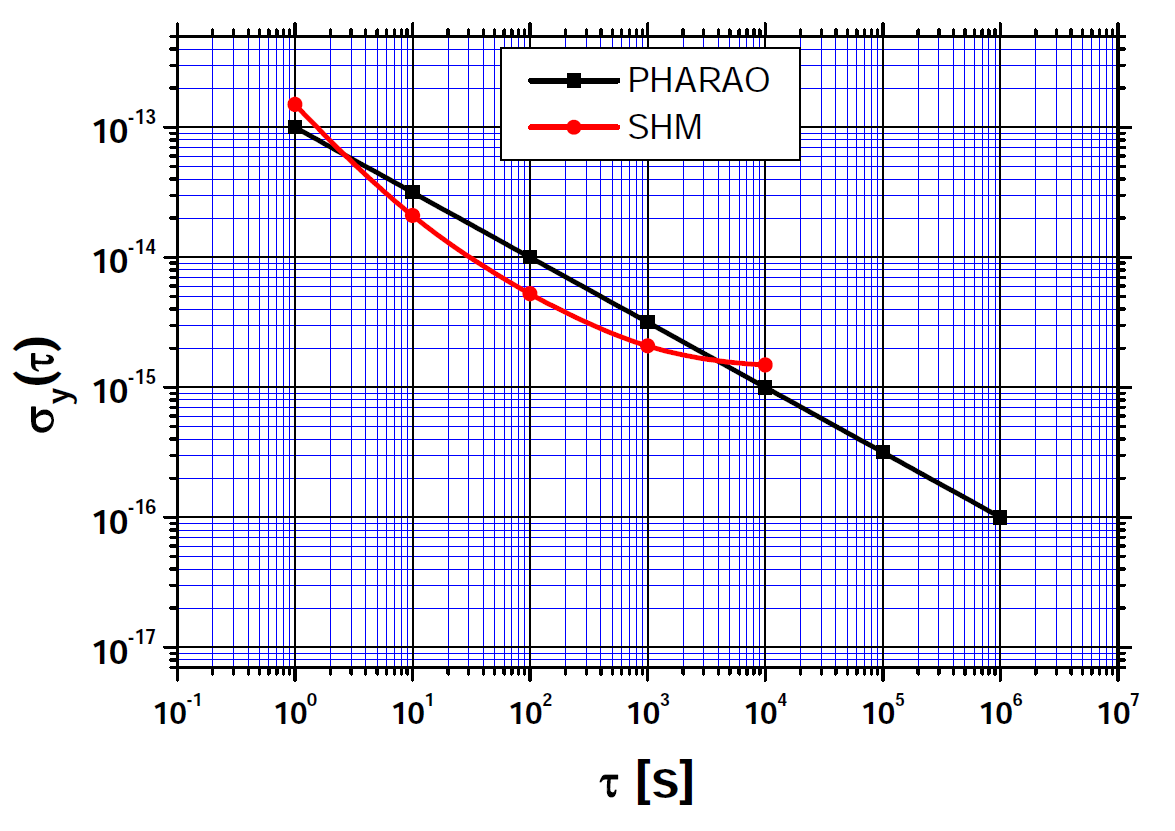}
        \caption{\label{f:sigma_y}PHARAO (Cesium clock) and SHM (hydrogen
    maser) expected performances in Allan deviation}
        
    \end{minipage} \end{figure}

The clock ensemble relative frequency stability (expressed in Allan deviation,
ADEV, see \textsl{e.g.}~\cite{Allan1966}) should be better than $\sigma_y =
10^{-13} \cdot \tau^{-1/2}$, which corresponds to $3\times10^{-16}$ after one
day of integration (see figure \ref{f:sigma_y}); time deviation (TDEV) should
be better than $2.1\times10^{-14}\cdot \tau^{1/2}$, which corresponds to 12~ps
after one day of integration (see figure \ref{f:sigma_x}). The fractional frequency uncertainty of the clock ensemble should be around $10^{-16}$. As the
expected gravitational frequency shift between ground and space clocks for an ISS altitude of 400~km is around $4\times10^{-11}$, this is coherent with a measurement of the gravitational redshift at the $3\times10^{-6}$ level.

The main scientific objectives of this mission are~\cite{ACES2009,Laurent2015}:

\begin{itemize} 
    \item To demonstrate the high performance of the atomic clocks
        ensemble in the space environment and the ability to achieve high
        stability on space-ground time and frequency transfer.  
    \item To compare
        ground clocks at high resolution on a worldwide basis using a link in
        the microwave domain. In common view mode, the link stability should
        reach around 0.3~ps after 300~s of integration; in non-common view
        mode, it should reach a stability of around 7~ps after 1~day of
        integration (see figure \ref{f:sigma_x}).  
    \item To perform equivalence
        principle tests. It will be possible to test the gravitational
        redshift with unprecedented accuracy, to carry out novel tests of
        Lorentz invariance and to contribute to searches for possible
        variations of fundamental constants.
\end{itemize}

Besides these primary objectives, several secondary objectives can be found
in~\cite{LC-001}. For example, comparing distant ground clocks at the
$10^{-17}$ uncertainty level in fractional frequency allows the determination
of the potential difference between their locations via the gravitational
redshift with an uncertainty of about 1 m$^2$/s$^2$, corresponding to an
uncertainty in height of about 10 cm, a method of geodesy referred to as
chronometric geodesy \cite{Vermeer1983,Delva2013,Delva2017b,Lion2017}.

These science objectives are directly linked to the MWL performance. For example, the short term ($\leq$ 300~s) stability of the MWL directly affects the test of Lorentz invariance \cite{Wolf1995,Delva2017a}, or the comparison of ground clocks at the $10^{-17}$ level, which in turn is required to efficiently contribute to searches for the variation of fundamental constants \cite{ACES2009}. The test of the gravitational redshift will likely be limited by the long term performance of the MWL and the on-board clocks, but ensuring adequate short term performance of the MWL is essential as well, e.g. for an efficient evaluation of systematic effects.

In this article we describe the operation of the MWL of ACES, with particular
emphasis on the theoretical description of the observables and the more general
theory of one-way and two way links for time and frequency transfer in free
space (section~\ref{sec2}). We show how scientific products (the main one being
the clock desynchronization) are obtained from the MWL measurements (section
\ref{sec3}) and give some details about our simulation and data analysis
software and the tests performed on their overall performance (section
\ref{sec4}). Finally we present the results of numerical simulations and
analysis concerning the impact of orbit determination errors on the simulated
clock desynchronization, which confirm the analytical results of
\cite{duchayne2009} (section \ref{sec5}).

\section{The ACES/PHARAO experiment} \label{sec2}

\subsection{General description}

The detailed setup of the ACES/PHARAO experiment is described in
\cite{ACES2009,Laurent2015}. The payload, a box with a volume of about 1~m$^3$,
will be launched using a Space X rocket in 2018, and attached to a balcony of
the European module, Columbus, on the International Space Station (ISS). The
planned mission duration is 18 months, with a potential extension to 3 years.
The payload contains a caesium cold atom clock (PHARAO) and a hydrogen maser
(SHM) and enables time and frequency comparison with ground clocks thanks to a
dedicated microwave link (a laser link will also be available but is out of
scope of this paper).

One peculiarity of this mission is that the links are integral part of the
measurement process : ground terminals (GTs) do not act as simple
telemetry/telecommand devices, their performance and calibration being a
possible limiting factor to the scientific outcome of the mission. They have
been designed and built especially for the ACES mission under ESA supervision.
It is currently foreseen that 9 of those Ground Terminals will be operational :
some will move during ACES flight, while others will remain fixed at several
metrology institutes.

Institutes hosting a GT are asked to provide a time reference, \textsl{i.e.} a
100 MHz sine plus a 1 PPS signal. This may be based on a local realisation of
UTC, or any stable signal whose relation to UTC is known (the time difference
should be known to better than the $\mu$s level). On board the ISS, a similar
100 MHz + 1 PPS combination is derived from PHARAO (the caesium beam clock) and
SHM (the active hydrogen maser) and fed to the MWL Flight Segment.

The GT hosts a directional antenna and a motorized alt-az mount, which tracks
the ISS whenever it is in sight. The flight segment's antenna is a phased array
antenna. Once the signal is locked, the link compares continuously (12.5 Hz
sampling rate) the ground signal to the space signal. The comparison ends when
the ISS goes below a given elevation, with passes lasting about 600 seconds at
most. For a given GT, such passes will occur at most 5 times per day, separated
by at least one orbital period (about 90 minutes).

\subsection{MWL Measurement Principle} \label{sec:MWL} The microwave link (MWL)
will be used for space-ground time and frequency transfer. A time transfer is
the ability to synchronize distant clocks, \textsl{i.e.} determine the
difference of their displayed time for a given coordinate time. The choice of
time coordinate defines the notion of simultaneity, which is only
conventional~\cite{Petit2014,Petit1994,Klioner1992}. A frequency transfer is
the ability to syntonize distant clocks, \textsl{i.e.} determine the difference
of clock frequencies for a given coordinate time.

The MWL is composed of three signals of different frequencies: one Ku-band
uplink at frequency $\simeq 13.5$~GHz, and two downlinks at $\simeq 14.7$~GHz
and 2.2~GHz (Ku and S-band, respectively).  Measurements are done on the
carrier itself and a pseudo-random noise (PRN) code that modulates the carrier
at 100~Mchip/s. In the following we will refer to those measurements as
``carrier'' and ``code'' observables and data. The link is asynchronous,
\textsl{i.e.} the uplink is independent of the downlink. Measurements are
provided at 80~ms intervals and can be interpolated in order to choose any
particular configuration, \textsl{e.g.} emission of the downlink signal
simultaneously with reception of the uplink signal at the spacecraft antenna
(so-called $\Lambda$ configuration, see \textsl{e.g.}~\cite{duchayne2009}).

Observables are similar to the ones of Global Navigation Satellite Systems
(GNSS). The emitted electromagnetic signal is locked to the clock signal of the
emitter, and the time of arrival of this electromagnetic signal is compared to
the clock signal generated by the receiver. Therefore the basic one-way
observable is a \emph{pseudo-time-of-flight} (PToF), which contains information
about both the time-of-flight of the signal and the difference of the times
given by the receiver and the emitter clocks. In GNSS it is more usual to talk
about \emph{pseudo-ranges}, which is the same as the PToF multiplied by $c$,
the velocity of light in vacuum. Note that the code has a frequency
of 100~MHz which leads to an accuracy for the time and frequency transfer worse
than the goal accuracy of the experiment. The carrier observable is therefore needed to reach the specifications, and as in GNSS we use the code to lift the carrier's cycle ambiguity.

The MWL data analysis main goal is to calculate from the PToF the
desynchronization between the ground and the space clocks. Therefore this
requires a model of the space-time geometry, a convention for defining the
simultaneity of events, as well as a model for the propagation of the signal,
which takes into account atmospheric and instrumental delays. The positions of
the ground and space stations have to be known to some level of accuracy in
order to calculate the time-of-flight. However when combining the downlink and
the uplink observables to extract clock desynchronization the range cancels to
first order. This is the main interest of two-way time and frequency transfer.
Moreover, the ionospheric delay is dispersive and it can be disentangled from
the clock desynchronization by analysing the two downlink observables, which
rely on signals at different frequencies and therefore have different
ionospheric delays.

\subsection{Spatio-temporal reference systems} \label{sec:transfo}

We are using the conventions adopted in 2000 and 2006 by the International
Astronomical Union as described in \cite{2003AJ....126.2687S}. It means that we
consider the Geocentric Celestial Reference System (GCRS) as our basic
reference frame where we calculate clock desynchronization. Consequently all
quantities are expressed with respect to the International Celestial Reference
frame (ICRF) and the Geocentric Coordinate Time (TCG) for the data analysis and
simulations. Final scientific products will be distributed in UTC timescale for
convenience.

First, we consider the positions of ground station and the ISS orbitography,
whose center of mass state vector (positions and velocities) are usually
distributed in SP3 files~\cite{1993pstn.conf..177R}. Usually, all these
quantities are expressed related to the International Terrestrial Reference
Frame (ITRF), which is body-fixed to the rotating Earth at one defining date,
as for example in 2008 and 2014~\cite{2011JGeod..85..457A,
2016JGRB..121.6109A}. It is then necessary to transform these state vectors
from ITRF to ICRF and it is done by using routines from SOFA package, developed
by IAU~\cite{SOFA:2016-05-03}. This requires the Earth Orientation Parameters
series (precession, nutation and polar motion) distributed by the International
Earth Rotation and Reference Systems Service (IERS).

At this point the position of the center of mass of the ISS is known in the
ICRF. We still need to find the position of the PHARAO microwave cavity and the
Ku and S-band antennas : for this we need to know the vector from the center of
mass to those elements, which is known and fixed in the Space Station
Coordinate System (SSACS). The orientation of SSACS with respect to ICRF is
given by a set of quaternions, the ``attitude'' of the station. From those
quaternions we derive the additional vector needed to calculate the position of
the cavity and antennas in the ICRF, thus producing distinct orbitography for
each.

\section{The Micro-Wave Link} \label{sec3}

\label{sec:simu}

In the following we give a formal description of one-way and two-way links for
code observables. The principle for carrier observables is the same except that
periods cannot be identified, leading to a phase ambiguity.

We suppose here that all clocks are perfect, \textsl{i.e.} their displayed time
is exactly their proper time. Proper time $\tau$ is given in a metric theory of
gravity by relation: \begin{eqnarray} \label{pr_time} c^2 \textrm{d} \tau^2 = -
    g_{\alpha \beta} \textrm{d} x^\alpha \textrm{d} x^\beta , \end{eqnarray}
where $g_{\alpha \beta}$ is the metric, $c$ the velocity of light, $\{ x^\alpha
\}$ the coordinates in the chosen reference system and Einstein summation rule
is used. We use superscript $A$ on proper times $\tau$ for clock $A$ and we
express proper time as a function of coordinate time $t$:
$\tau^A\equiv\tau^A(t)$. Moreover, we use the notation $[.]$, which is the
coordinate~/~proper time transformation obtained from equation \eqref{pr_time},
and $T_{ij} = t_j-t_i$ for coordinate time intervals\footnote{\textsl{e.g.}
$[T_{12}]^A$ is the transformation of coordinate time interval $(t_2-t_1)$ to
proper time of clock $A$, and $[\Delta \tau^A]^t$ is the transformation of
proper time interval $\Delta \tau^A$ of clock $A$ to coordinate time $t$.}.

\subsection{From Pseudo-Time of Flight to scientific products} \label{sec:1way}

\begin{figure}[ht] \centerline{ \subfigure[Sequence of events]{
        \resizebox{\linewidth}{!} {
            \includegraphics[width=\linewidth]{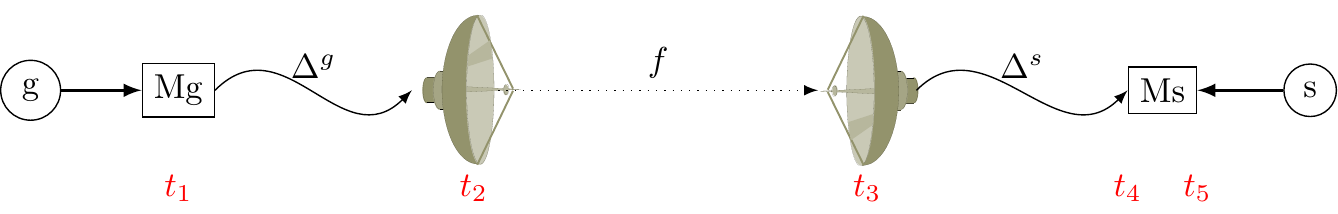} } \label{f:1way1} } }
            \centerline{ \subfigure[Space-time diagram]{
                \resizebox{0.8\linewidth}{!} {
                    \includegraphics[width=\linewidth]{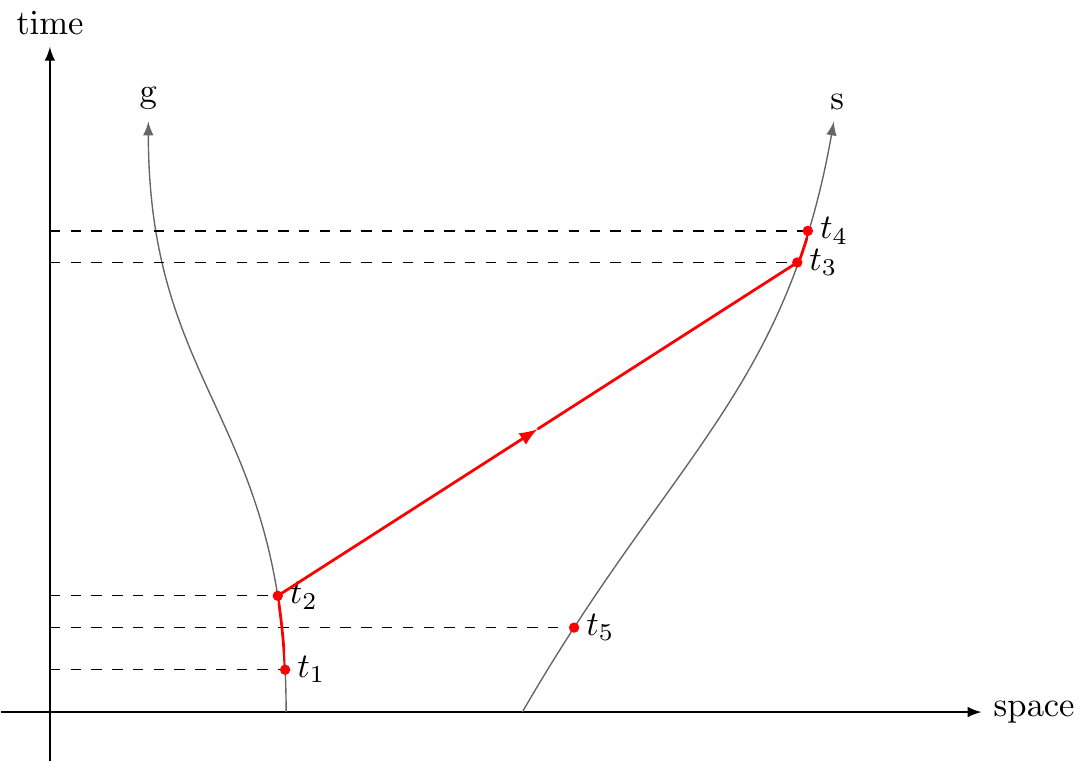} }
                    \label{f:1way2} } } \caption{Schematic representation of
                    the one-way link.} \label{f:1way} \end{figure}

Let's consider a one-way link between a ground and a space clock represented
respectively by superscripts $g$ and $s$. The sequence of events is illustrated
on figure \ref{f:1way}. At time coordinate $t_1$, clock $g$ displays time
$\tau^g(t_1)$ and modem Mg produces a code $C^1$. This code modulates a
sinusoidal signal of frequency $f$ and is sent at coordinate time $t_2$ by
antenna $g$. The time delay between the code production and its transmission by
antenna $g$ is $\Delta^g = [T_{12}]^g$, expressed in local frame of clock~$g$.
Antenna $s$ receives signal $C^1$ at coordinate time $t_3$, and transmits it to
modem Ms and clock~$s$ which receives it at coordinate time $t_4$, with a delay
$\Delta^s = [T_{34}]^s$, expressed in local frame of clock~$s$. 

The codes produced by the ground and the space clocks are the same, meaning
that for the same proper time displayed by the clocks the same piece of code is
produced. This is an unambiguous way to synchronize the signal to the proper
time of the clocks. We implicitly define the coordinate time $t_5$ when clock
$s$ displayed the same code $C^1$ as encoded in the received signal:
\begin{equation} \label{e:1way1} \tau^s (t_5) = \tau^g (t_1) \ .
\end{equation} The pseudo-time-of-flight $\Delta \tau^s$ (PToF, see
section~\ref{sec:MWL}) given by the receiver~$s$ is then defined as:
\begin{equation} \label{e:1way2} \Delta \tau^s ( \tau^s (t_4) ) \equiv \tau^s
    (t_5) - \tau^s (t_4) .  \end{equation} The PToF is dated with proper time
of clock $s$ when this clock receives code $C^1$ from antenna $s$. It can be
interpreted as the difference between the proper time of production of code
$C^1$ by clock $s$, and the proper time of reception of the same code $C^1$
sent by clock $g$, all expressed in proper time of clock $s$.

Then the desynchronization $\delta$ between clocks $g$ and $s$ is written in a
hypersurface characterized by coordinate time $t=\mathrm{constant}$. From
equations \eqref{e:1way1}-\eqref{e:1way2} it is straightforward to calculate
the desynchronization at coordinate time~$t_4$:
\begin{equation} \label{e:desync} \delta (t_4) \equiv \tau^s (t_4) - \tau^g
    (t_4) = - \Delta \tau^s \left( \tau^s (t_4) \right) - \left[ T_{23} +
        \left[ \Delta^g + \Delta^s \right]^t \right]^g \end{equation} Similar
    formulas can be obtained for the desynchronization at coordinate
    times~$t_1$ and~$t_5$. This expression has been obtained for the uplink,
    from ground to space. The desynchronization expression with downlink
    observables may be obtained by exchanging $g$ and $s$ in equation
    \eqref{e:desync}.

\begin{figure}[ht!] \subfigure[Uplink sequence of events]{
        \resizebox{\linewidth}{!} {
            \includegraphics[width=\linewidth]{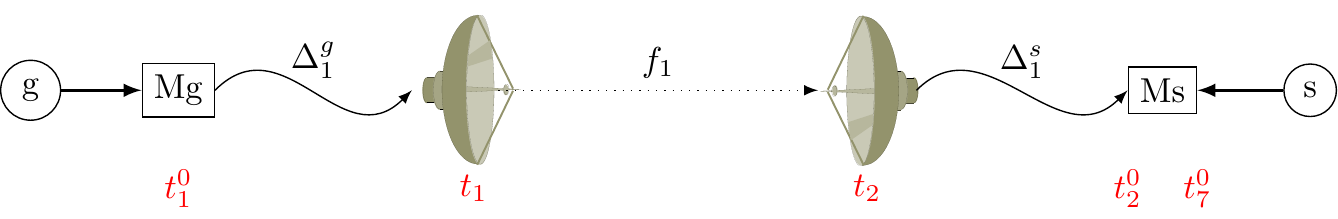}} \label{f:2way1}}
            \subfigure[Downlink sequence of events]{ \resizebox{\linewidth}{!}
            { \includegraphics[width=\linewidth]{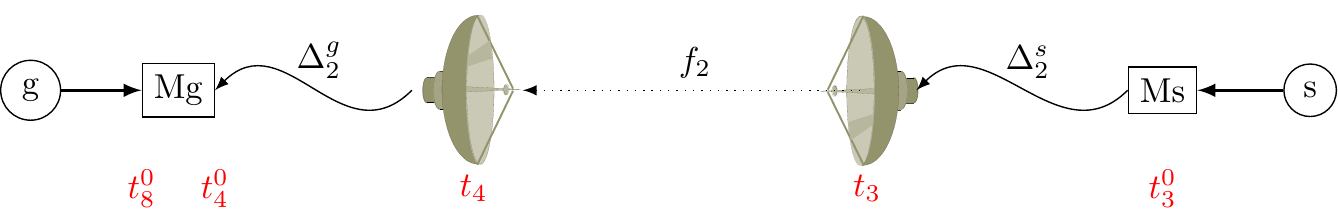}} \label{f:2way2}}
            \centerline{\subfigure[Space-time diagram]{
                \resizebox{0.8\linewidth}{!} {
                    \includegraphics[width=\linewidth]{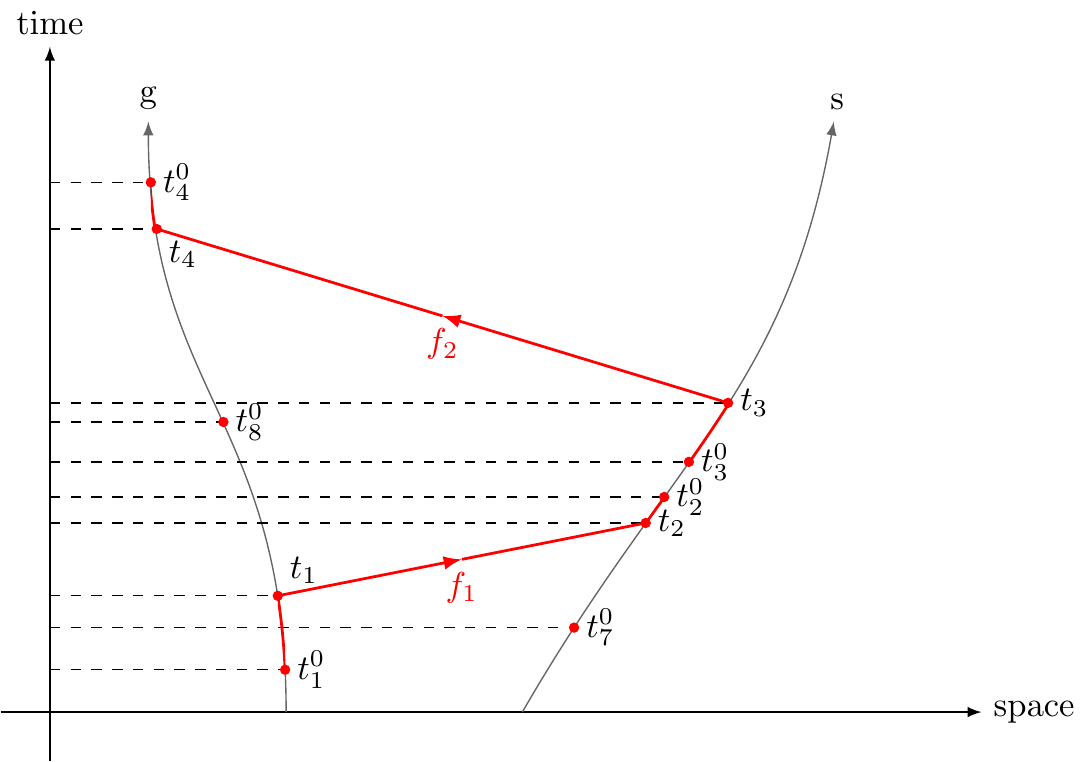}}
                    \label{f:2way3} }} \caption{Schematic representation of the
                    two-way link.} \label{f:2way} \end{figure}

Let's consider now a link which is composed of two one-way links, between a
ground and a space clock represented respectively by subscript $g$ and $s$.
The two sequences of events are illustrated on figure ~\ref{f:2way}. The uplink
(from ground to space) has a frequency $f_1$ and is represented by coordinate
time sequence $(t_1^0, t_1, t_2, t_2^0, t_7^0)$ (figure \ref{f:2way1}). This
link is defined with the relation:
\begin{equation} \tau^g (t_1^0) = \tau^s (t_7^0) .  \end{equation}
The downlink has a frequency $f_2$ and is represented by coordinate time
sequence $(t_3^0, t_3, t_4, t_4^0, t_8^0)$ (figure \ref{f:1way2}). This link is
defined with the relation:
\begin{equation} \tau^s (t_3^0) = \tau^g (t_8^0) .  \end{equation}

In the ACES/PHARAO mission we use the so-called $\Lambda$ configuration. This
configuration minimizes the error coming from the uncertainty on ISS
orbitography (in~\cite{duchayne2009} it has been shown that in this
configuration the requirement on ISS orbitography accuracy is around 10~m).
The $\Lambda$ configuration is defined by $t_2=t_3$, \textsl{i.e.} code $C^2$
is sent at antenna~$s$ when code $C^1$ is received at this antenna. This
configuration is obtained by interpolating the observables in the data analysis
software. Then it can be shown from equation~\eqref{e:desync} that the
desynchronization between clocks $g$ and $s$ at coordinate time~$t_2$ is:

\begin{equation} \label{eq:lambda} \delta (t_2) \equiv \tau^s (t_2) - \tau^g
    (t_2) = \frac{1}{2} \left( \Dtc^g (t_4^0) - \Dtc^s (t_2^0) + \left[ T_{34}
    - T_{12} \right]^g \right) .  \end{equation} where we introduced the
\emph{corrected} observables $\Dtc^g$ and $\Dtc^s$: \begin{equation}
    \label{e:mod} \begin{array}{lcl} \Dtc^g (t_4^0) &=& \Delta \tau^g \left(
        \tau^g (t_4^0) \right) + \Delta^g_2 + \Delta^s_2\\[0.1in] \Dtc^s
        (t_2^0) &=& \Delta \tau^s \left( \tau^s (t_2^0) \right) + \Delta^g_1 +
        \Delta^s_1.  \end{array} \end{equation}

In equation~\eqref{eq:lambda} remains one transformation from coordinate to
proper time. We know that $T_{12} \sim T_{34} \sim 1$~ms. During this time
interval the gravitational potential and velocity of the ground station can be
considered to be constant, leading to: \begin{equation} \label{eq:approx}
    \left[ T_{34} - T_{12} \right]^g = ( 1 - \epsilon_g (t_2) ) \left( T_{34} -
    T_{12} \right) , \end{equation} where \begin{equation} \nonumber \epsilon_g
    (t) = \dfrac{G M}{r_g(t) c^2} + \dfrac{v_g^2(t)}{2 c^2} , \end{equation}
$M$ is the Earth mass, $r_g(t)$ and $v_g(t)$ are the ICRS radial coordinate and
the coordinate velocity of the ground clock at coordinate time $t$. 

Orders of magnitude of these corrective terms are: \begin{eqnarray} \dfrac{G
    M}{r_g c^2} T_{34} &\sim& 0.6 \ \textrm{ps} \nonumber \\ \dfrac{v_g^2}{2
    c^2} T_{34} &\sim& 0.002 \ \textrm{ps}\nonumber \end{eqnarray} The
gravitational term is just at the limit of the required accuracy. The velocity
term is well below, so we can neglect it.  The final formula for
desynchronisation is then: \begin{equation} \label{e:desyncfin} \delta(t_2) =
    \frac{1}{2} \left[ \Dtc^g (t_4^0) - \Dtc^s (t_2^0) + \left( 1-\frac{G
    M}{r_g(t_2) c^2} \right) \left( T_{34} - T_{12} \right) \right] .
\end{equation}

In the $\Lambda$ configuration we suppose that $T_{23}=0$. However, this will
never be exactly~0 and it will be known with an accuracy $\delta T_{23}$.  This
will add a supplementary delay $\delta (\tau^s - \tau^g)$ to
desynchronization~\eqref{eq:lambda}: \begin{equation} \delta (\tau^s - \tau^g)
    (t_2) = \left( \epsilon_g (t_2) - \epsilon_s (t_2) \right) \delta T_{23}
\end{equation} Orders of magnitude are: \begin{eqnarray} \dfrac{G M}{c^2}
    \left( \dfrac{1}{r_g} - \dfrac{1}{r_s} \right) &\sim& 2.8\times10^{-11}
    \nonumber\\ \dfrac{v_g^2-v_s^2}{2 c^2} &\sim& -3.3\times10^{-10} \nonumber
\end{eqnarray} With the required accuracy on the MWL, $|\delta  (\tau^s -
\tau^g)| \lesssim 0.3$~ps, we deduce the following constraint on $\delta
T_{23}$: \begin{equation} \nonumber \delta T_{23} \lesssim 0.9 \ \textrm{ms}.
\end{equation} This constraint is much less constraining than the one coming
from orbitography, which is $\delta T_{23} \lesssim 1~\mu\textrm{s}$
(see~\cite{duchayne2009}).

\subsubsection{Atmospheric delays}

The downlink is composed of two one-way links of frequencies $f_2$ and $f_3$,
represented respectively by coordinate time sequence $(t_3^0, t_3, t_4, t_4^0,
t_8^0)$ and $(t_5^0, t_5, t_6, t_6^0, t_9^0)$. These two links are affected by
a ionospheric delay that depends on their respective frequencies, whereas the
tropospheric delay does not depend on the link frequency. Dispersive
troposphere effects can marginally reach the required uncertainties but are
neglected at this stage. They can be taken into account using a global
model~\cite{hobiger2013correction}. We write: \begin{eqnarray} T_{12} & =
    \frac{R_{21}}{c} + \Delta^{\mathrm{iono}}_{12} (f_1) +
    \Delta^{\mathrm{tropo}}_{12} + \Delta^{\mathrm{Shapiro}}_{21}
    \label{eq:T12}\\ T_{34} & = \frac{R_{34}}{c} + \Delta^{\mathrm{iono}}_{34}
    (f_2) + \Delta^{\mathrm{tropo}}_{34} + \Delta^{\mathrm{Shapiro}}_{34}
    \label{eq:T34}\\ T_{56} & = \frac{R_{56}}{c} + \Delta^{\mathrm{iono}}_{56}
    (f_3) + \Delta^{\mathrm{tropo}}_{56} + \Delta^{\mathrm{Shapiro}}_{56}
    \label{eq:T56} \end{eqnarray} where $R_{ij} = |\vxR (t_j) - \vxT (t_i) |$
is the range, $\vxT$ and $\vxR$ are respectively position vectors of space and
ground antennas, $\rT = |\vxT|$, $\rR = |\vxR|$ and:
\begin{equation} \Delta^\text{Shapiro}_{ij} = \dfrac{2GM}{c^3} \ln \left(
    \dfrac{\rT(t_i)+\rR(t_j)+R_{ij}}{\rT(t_i)+\rR(t_j)-R_{ij}} \right) +
    \Ol(c^{-4}) \end{equation}

Ionospheric and tropospheric delays are around or below 100~ns, whereas Shapiro
delay (term in $c^{-3}$) is below 10~ps for the ACES/PHARAO mission
(see~\cite{duchayne:Thesis:2003} and \ref{sec:appendix}).

\subsubsection{Tropospheric delay and range} \label{sec:tropo} By adding ground
and space observables of links $f_1$ and $f_2$ we obtain: \begin{equation}
    \nonumber \begin{array}{l} \Delta \tau^s ( \tau^s(t_2^0) ) + \Delta \tau^g
        ( \tau^g(t_4^0) ) + \Delta^g_1 + \Delta^g_2 \\[0.1in] + \left[ \left[
        \Delta^s_1 + \Delta^s_2 \right]^t \right]^g = \left[ T_{23}^0 \right]^s
        - \left[ T_{23}^0 \right]^g - \left[ T_{12} + T_{34} \right]^g
    \end{array} \end{equation} As in the previous section, it can be shown that
$\left[ T_{23}^0 \right]^s - \left[ T_{23}^0 \right]^g \simeq 0$ if $T_{23}^0$
is known with an accuracy $\delta T_{23}^0 \lesssim 0.9$~ms. We neglect the
coordinate to proper time transformations for delays  $\Delta^s$ and obtain
\begin{equation} \nonumber T_{12} + T_{34} = - \left( 1 + \dfrac{G M}{r_g (t_2)
    c^2} \right) \left( \Dtc^s (t_2^0) + \Dtc^g (t_4^0) \right) ,
\end{equation} Then, from equations \eqref{eq:T12}-\eqref{eq:T34} we obtain:
\begin{eqnarray} \nonumber \dfrac{R_{21} + R_{34}}{c} &=& - \left( 1 + \dfrac{G
    M}{r_g (t_2) c^2} \right) \left( \Dtc^s (t_2^0) + \Dtc^g (t_4^0) \right)\\
    &&- \left( \Delta^{\text{iono}}_{12} (f_1) + \Delta^{\text{iono}}_{34}
    (f_2) + \Delta^{\text{tropo}}_{12}  + \Delta^{\text{tropo}}_{34} \right)
    \label{eq:degen} \\ && \nonumber - \left( \Delta^{\text{Shap}}_{21} +
    \Delta^{\text{Shap}}_{34} \right) \end{eqnarray}
where Shapiro delay is added for completeness; practically it should be
negligible compared to the error of the tropospheric model.  The ionospheric
delays are obtained from the two downlink observables (see \ref{sec:appendix}),
but as can be seen from equation \eqref{eq:degen} the range and tropospheric
delays are degenerate.  Range can be calculated with a model for tropospheric
delay, and tropospheric delay can be calculated from an estimation of range.

\subsection{From raw data to Pseudo-Time of Flight} \label{sec:raw}

Initial algorithms development have been made assuming that the microwave link
data would be similar to pseudo-time of flight (PToF), as issued by GNSS
receivers. In the course of hardware development, it appeared that a rather
different approach would be used by the manufacturer, based on local signal
beatnotes and counter measurements. In a first and independent
``preprocessing'' step our software transforms those raw measurements into PToF
that are then used in further processing.

\begin{figure}[ht!] \centering \includegraphics[width=\linewidth]{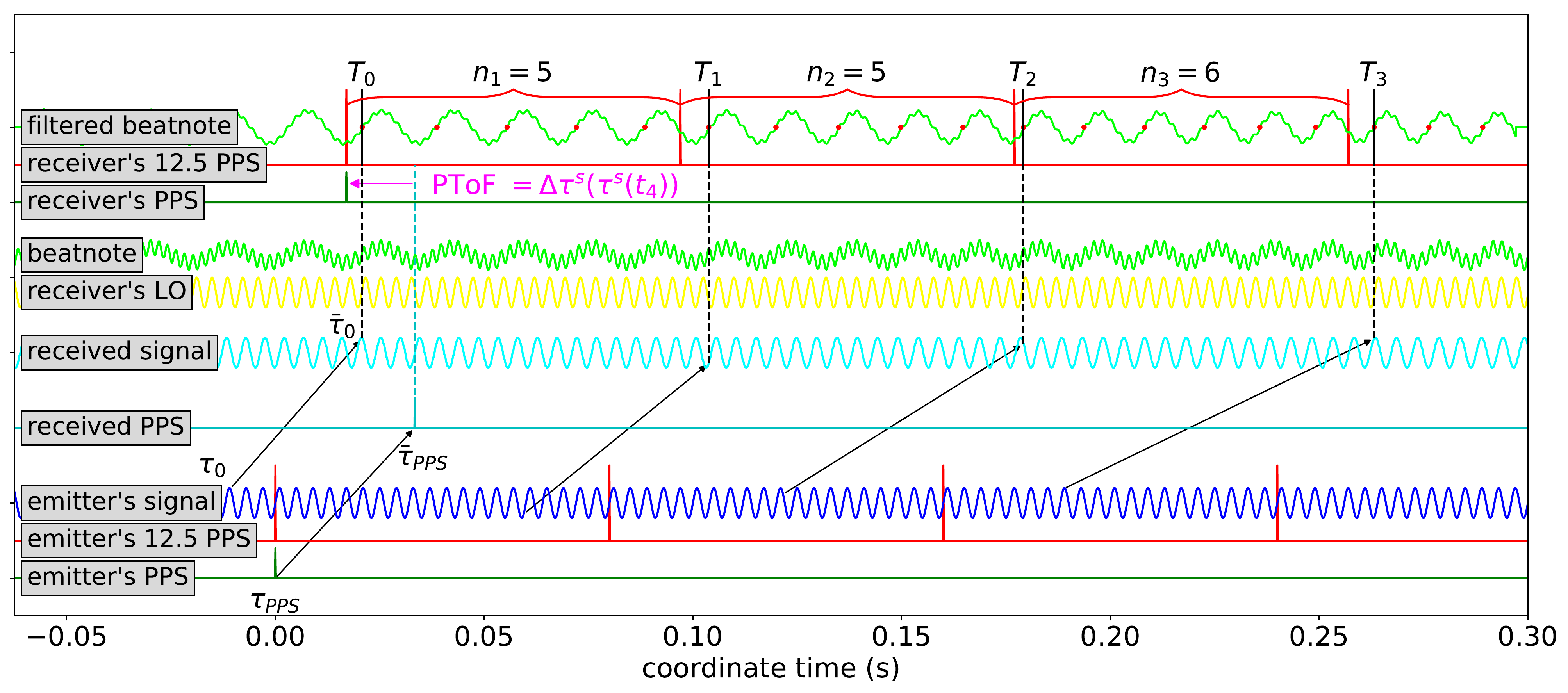}
    \caption{Schematic representation of the measurements provided by the
        hardware: both emitter's and receiver's signals are represented against
        the coordinate time scale. 1 PPS (dark green) and 12.5 PPS (red) pulses
        for the emitter and the receiver are represented. The emitter's carrier
        is represented by the blue sine at the bottom. Propagation delays are
        represented by black arrows. Incoming PPS and carrier are affected by
        this (variable) delay and appear in cyan. Local oscillator (yellow) is
        mixed with incoming carrier to give the green beatnote, which is then
        filtered. Red dots on the filtered beatnote indicate zero crossings on
        ascending edge. $n_m$ is the number of red dots between two 12.5 PPS
        pulses.  $T_{m-1}$ is the proper time of the first red dot within the
        same interval.  Signal frequency is much lower than in reality. Doppler
        effect is strongly magnified in order to show the variation of $n_m$.
    Finally, the observable we want to determine (PToF) is represented as a
purple arrow.} \label{fig:beatnote} \end{figure}

Detailed description of the algorithms is out-of-scope of this article, but an
overview may be found in figure \ref{fig:beatnote}. It can roughly be read as a
chronogram, with several signals being superposed on it :

\begin{itemize} \item The 3 bottom lines represent a sinusoidal signal, a 1 PPS
            and a 12.5 PPS that delimitates 80 ms intervals. These 3 signals
            are generated at the emitter's level, and are locked to the local
            clock signal (they are the best realization of the clock proper
            time): for simplicity, we represented the first PPS of the emitter
            at the origin of coordinate time.

    \item The ``received signal'' corresponds to the emitter's signal as
        received by the remote station: the black arrows represent the
        propagation delay for several events, which changes throughout the
        pass, and produces a classical Doppler effect.

    \item The upper part of the graph represents what happens at the receiver's
        level : 1 and 12.5 local PPS are generated, together with a local sine
        signal (``receiver's Local Oscillator''). Mixing the received signal
        and the local oscillator generates a beatnote, which is filtered
        through a low-pass filter.

    \item Then the measurement takes place : for each 80 ms period, the device
        counts the number of ascending zero-crossings of the filtered beatnote,
        $n_m$, and the date of the first zero-crossing, $T_m$. $n_m$ variations
        are mostly due to Doppler effect on received signal. Note that
        measurement occurs at the first zero crossing, so it is not natively
        aligned with 80 ms interval.  \end{itemize}

From $(T_m, n_m)$ couples, we have to recover the time difference represented
in purple: $\Delta\tau^s(\tau^s(t_4))$, which is a PToF : for a given proper
time, the main contributions to this value are the desynchronisation between
the two clocks, minus the propagation delay.

\subsection{Code and carrier ambiguities}

\label{sec:cocaamb}

As explained in the previous section, we get from the receiver modem (space and
ground) three kinds of files: pulse (PPS arrival times), code ($T_m$, $n_m$)
and carrier ($T_m$, $n_m$). Code is in principle unambiguous (within the
code's length) and chip counters are implemented to ensure that. However, for historical and technical reasons, we have chosen to not use the chip counters, but instead treat the code the same way as the carrier i.e. to work with a code ambiguity corresponding to a 100~MHz cycle (one chip length), and, as for the carrier, lift that ambiguity using a coarser measurement,
namely the pulse data. From it, we deduce coarse times of emission and
reception of the signal, roughly every 1 second, such that we can directly
calculate the PToF. However we are limited in accuracy by the quantization of
the time counter ($\approx$ 10 ns) that registers the PPS arrival times and the
$T_m$ value, well above the targeted accuracy.  This pulse observable will be
used to remove the ambiguity on the code, then on the carrier observables. In
this section we explain one method allowing to find the code and carrier
ambiguities.

Let's consider the one-way link described in section~\ref{sec:1way}. The signal
is sent at $\tau\equiv\tau^g(t_1)$ in the local time of the emitter, and
received at $\bar\tau\equiv\tau^s(t_4)$ in the local time of the receiver,
neglecting here, for simplicity, instrumental and ionospheric delays. Let
$\Phi\LO$ be the phase of the receiver local oscillator, while $\Phi\Rx$ and
$\Phi\Tx$ are respectively the phase of the received and emitted signal (see
figure~\ref{fig:beatnote}). Then: \begin{eqnarray} \Phi\LO (\bar\tau) =
    \omega\LO \bar\tau + \phi\LO^0 \label{eq:phiLO}\\ \Phi\Rx (\bar\tau) =
    \Phi\Tx (\tau) = \omega\Tx \tau + \phi\Tx^0 \label{eq:phiRx} \end{eqnarray}
where the origin of the phases $\phi\LO^0$ and $\phi\Tx^0$ are unknown. There
is a known relation between the local oscillator phase $\Phi\LO$, the phase of
the received signal $\Phi\Rx$ and the beatnote phase $\Phi\beat$. The angular
frequency of the code is higher than the angular frequency of the code local
oscillator, such that $\Phi\beat = + (\Phi\LO - \Phi\Rx)$ for the code signal,
while the angular frequency of the carrier is lower than the angular frequency
of the carrier local oscillator, such that $\Phi\beat = - (\Phi\LO - \Phi\Rx)$
for the carrier signal. Using equations~\eqref{eq:phiLO}-\eqref{eq:phiRx}, we
deduce: \begin{eqnarray} \omega\Tx \Delta\tau (\bar\tau) =  ( \omega\beatc
    \bar\tau - \Phi\beat (\bar\tau) ) + \phi\LO^0 - \phi\Tx^0
    \label{eq:dtau1}\\ \omega\LO \Delta\tau (\bar\tau) =  ( \omega\beatc \tau -
    \Phi\beat (\bar\tau) ) + \phi\LO^0 - \phi\Tx^0 \label{eq:dtau2}
\end{eqnarray} where $\Delta\tau (\bar\tau) = \tau - \bar\tau$ is the PToF, and
$\omega\beatc$ is the ``zero Doppler'' beatnote angular frequency,
\textsl{i.e.} $\omega\beatc = \pm (\omega\LO - \omega\Tx)$ ($+$ sign for code
and $-$ sign for carrier). Typical values of $\omega\beatc / (2\pi)$ are 195
kHz and 729 kHz for the code and carrier respectively. 

The ambiguity in the calculation of the PToF $\Delta\tau$ comes from the fact
that $(\phi\LO^0 - \phi\Tx^0)$ is unknown. Fortunately, the code and pulse
signals are designed such that $(\phi\LO^0 - \phi\Tx^0)/(2\pi)$ is an integer.
Moreover, the beatnote phase $\Phi\beat$ can be interpolated from the modem
code observables $(T_m,n_m)$ (see previous section) up to a multiple of
$(2\pi)$, and we call $\tilde\Phi\beat$ the interpolated beatnote. Therefore
one has: \begin{equation} \phi\LO^0 - \phi\Tx^0 - \Phi\beat (\bar\tau) = 2\pi k
    - \tilde\Phi\beat (\bar\tau) \end{equation} where $k$ is an unknown
integer, and we deduce from equations~\eqref{eq:dtau1}-\eqref{eq:dtau2} for the
pulse and the code: \begin{eqnarray} f\Tx\co \Delta\tau (\bar\tau) = f\beatc\co
    \bar\tau + k - \frac{\tilde\Phi\beat\co (\bar\tau)}{2\pi}
    \label{eq:dtau3}\\ f\LO\co \Delta\tau (\bar\tau) = f\beatc\co \tau + k -
    \frac{\tilde\Phi\beat\co (\bar\tau)}{2\pi} \label{eq:dtau4} \end{eqnarray}

Relation (\ref{eq:dtau4}) can be used with the pulse observables to find $k$,
because we know both $\tau$ and $\bar\tau$ for the pulse, and the 10~ns
accuracy of the pulse PToF $\Delta\tau$ can be shown to be sufficient after a
little averaging ($k$ remains constant during continuous lock of the signal
\textsl{e.g.} during a pass). Using the value of $k$ found with the pulse
observable we then remove the ambiguity on the code observable using
relation~(\ref{eq:dtau3}).

For the carrier signal the quantity $(\phi\LO^0 - \phi\Tx^0)/(2\pi)$ is not an
integer anymore. Therefore relation~(\ref{eq:dtau1}) can be written for the
carrier observable: \begin{equation} f\Tx\ca \Delta\tau (\bar\tau) = -
    f\beatc\ca \bar\tau + j + \frac{\tilde\Phi\beat\ca (\bar\tau)}{2\pi} +
    \frac{\phi\ca_0}{2\pi} \label{eq:dtau_ca} \end{equation} where $j$ is an
unknown integer and $0 \le \phi\ca_0 < 2\pi$ is an unknown phase origin. This
relation can be used to find both $j$ and $\phi\ca_0$: one has to interpolate
the absolute code series found earlier using relation~(\ref{eq:dtau3}), with an
error of the order of 20~ps. Therefore we can find $j$ for each pass, but the
determination of $\phi\ca_0$ will be limited by the accuracy of the absolute
code series. Then it is not possible to find a better determination of the
absolute PToF for one individual pass using the carrier. However, $\phi\ca_0$
is common to all passes; using several passes it is possible to average the
error done on its determination up to an error $\sim 20$~ps$/\sqrt{N}$, where
$N$ is the number of passes. It would then take around 1600 passes to attain
the targeted carrier accuracy which is of the order of 0.5~ps. Fortunately, for
most science objectives, we do not need this accuracy on the absolute PToF but
only on their differences, for which the term $\phi\ca_0$ cancels out.

\section{Data analysis and simulation} \label{sec4}

\subsection{Implementation of the simulation} 
\begin{figure}[ht!] \centering
    \includegraphics[width=\linewidth]{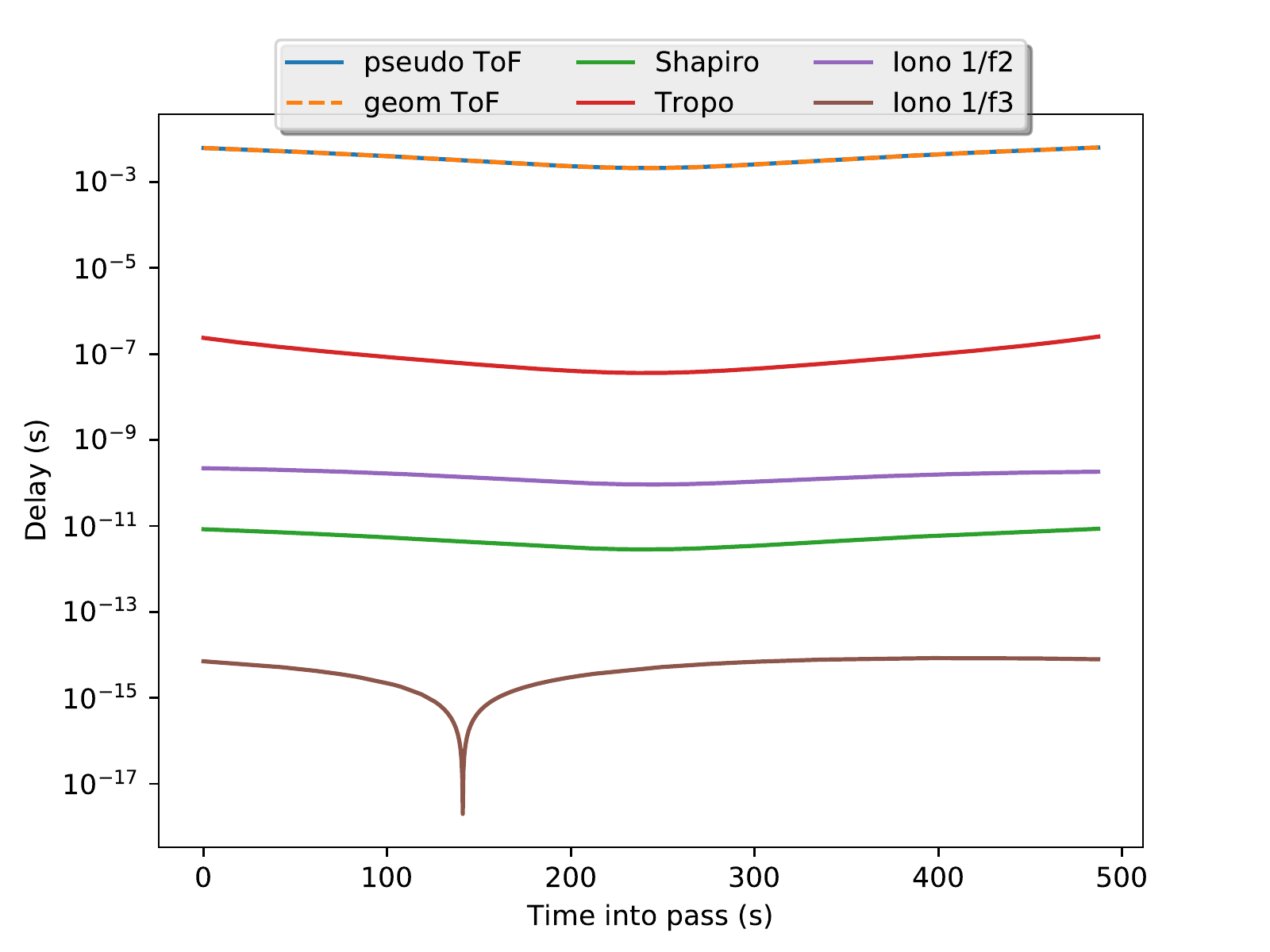} \caption{Contribution of
    atmospheric and Shapiro delays in the simulated time-of-flight of the $f_1$
signal for carrier observables (absolute values). Pseudo ToF is the sum of all
other components and is dominated by geometrical time of flight (curves are
superposed here). Ionospheric delay is split between its $1/f^2$ and $1/f^3$
terms, the latter being negligeable for Ku-band data but much larger for
S-band.} \label{f:tof} \end{figure}

In order to test the data analysis software, a simulation of the mission
observables has been written in Matlab ($\simeq4000$~lines). It is as
independent as possible from the data analysis software: different coding
language, algorithms, etc\ldots It takes as input orbitography of the ISS and
terrestrial coordinates of a Ground Station~(GS). The ISS trajectory is either
simulated (Kepler orbit) or a real trajectory read from an sp3 file
(\url{https://igscb.jpl.nasa.gov/igscb/data/format/sp3c.txt}). In the case of a
real trajectory for the ISS and for the GS, the coordinate transformations from
terrestrial to celestial frame are computed with the SOFA routines (see
section~\ref{sec:transfo}). For a given number of days the simulation then
calculates the times of passes for the GS corresponding to a given elevation
cutoff of the ISS.

From the trajectories and a model of the geo-potential the simulation
calculates the relation between proper times given by ISS and GS clocks and the
coordinate time. In an effort to keep numerical accuracy below the targeted
accuracy (timing error below 0.3~ps throughout one pass and between passes)
local timescales for all passes, linked to a global timescale, have been
introduced. The time-of-flight (ToF) of the signal has several contributions:
geometrical, ionospheric, tropospheric and Shapiro (see equations
\eqref{eq:T12}-\eqref{eq:T56}). The geometrical ToF is calculated numerically
by solving a two-point boundary value problem by means of the shooting
method~\cite{Miguel2007}. The ionospheric delay depends on the electron density
profile in the atmosphere (see equations \eqref{e:ionod1}-\eqref{e:ionod2}) and
on the magnetic field. The STEC $S$ can be chosen as constant or integrated
from a Chapman layer model~\cite{chapman1931} for a more realistic delay. The
magnetic field is a simple dipole model in the simulation, while we have chosen
a more sophisticated model for the data analysis in order to check the effect
of a mismodeling of the magnetic field. We can see no effect above the targeted
noise due to this difference of model. Finally, we use a Saastamoinen model in
order to calculate the tropospheric delay~\cite{saasta73}.

The Saastamoinen model is not really reliable at low elevation of ISS and the
dispersive part of troposphere has not been taken into account. Indeed, the
main scientific product of the ACES experiment is desynchronization, which does
not require a high precision tropospheric model to be calculated. The only
constraint is to be able to perform the $\Lambda$-configuration interpolation,
which requires an accuracy better than 30~m on the range. To this end the
Saastamoinen model is sufficient. However, in order to obtain the best accuracy
for the range as a secondary scientific product, it is necessary to disentangle
the geometric ToF from the tropospheric ToF (see section~\ref{sec:tropo}).
Independent studies have been performed~\cite{hobiger2013correction,
delva_tropospheric_2012} in order to study the best available model for the
tropospheric delay, including the dispersive part.

In figure \ref{f:tof} we have plotted different contributions included in the
ToF of signal $f_1\simeq 13.5$~GHz. Minimum elevation of ISS is taken as
10~degrees, and atmospheric parameters are temperature $T=298$~K, pressure
$p=1$~bar and water vapor pressure $e=0.5$~bar. A sinusoidal variation is added
to atmospheric parameters, with a period of 24~hours, and a Chapman layer model
is used to calculate the STEC.  Tropospheric delay is dominant in the ToF, with
a value of several 10~ns. Ionosphere is dispersive such that ionospheric delay
can be separated in two contributions: an effect that scales with $1/f^2$ and
one that scales with $1/f^3$ (see equations~\eqref{e:ionod1}-\eqref{e:ionod2}).
The second order contribution is around 1~ns and the third contribution around
0.1~ps, below mission accuracy. However these effects are much larger for the
$f_3=2.25$~GHz signal: around 40~ns for second order term and 20~ps for third
order term. Then third order terms cannot be ignored. The Shapiro delay of
several ps is slightly over the required accuracy, but, as already mentioned,
cancels in the two-way combination and is negligible with respect to
tropospheric model uncertainties in one-way or range measurements.

Finally, pseudo-time-of-flight (PToF) are calculated from the ToF and the
coordinate to proper time transformation for the GS and the ISS. Then PToF are
used to generate the modem observables $(T_m,n_m)$ (see section~\ref{sec:raw}).
This procedure is repeated six times (for the three signals, and for the code
and carrier) in order to produce the simulated observables files which are in
the same format as the operational files. Some features can be added optionally
such as signal multipath, non-null initial desynchronization of the clocks,
clock noise and orbitography errors. The simulation produces all blue boxes
(except calibration data) and red boxes shown in
figure~\ref{fig:global_flowchart} in order to test the data analysis software.

\subsection{Data analysis software} Software has been developped in Python3
(\url{https://www.python.org/}), relying on NumPy and SciPy \cite{numpy, scipy}
for the (moderately) heavy numerical calculations, matplotlib \cite{matplotlib}
being used for data graphing. This allows satisfactory performance while
keeping good readability and maintainability. The module has been encapsulated
as a Python package for easier deployment and version management on production
servers.

From the onset, this software has been designed as a pipeline, intended to run
as automatically as possible. However it is also possible to process individual
data, while picking which modules should be run (\textsl{e.g.} troposphere
delay, ionosphere delay, etc\ldots).

\begin{figure}[ht] \includegraphics[width=\linewidth]{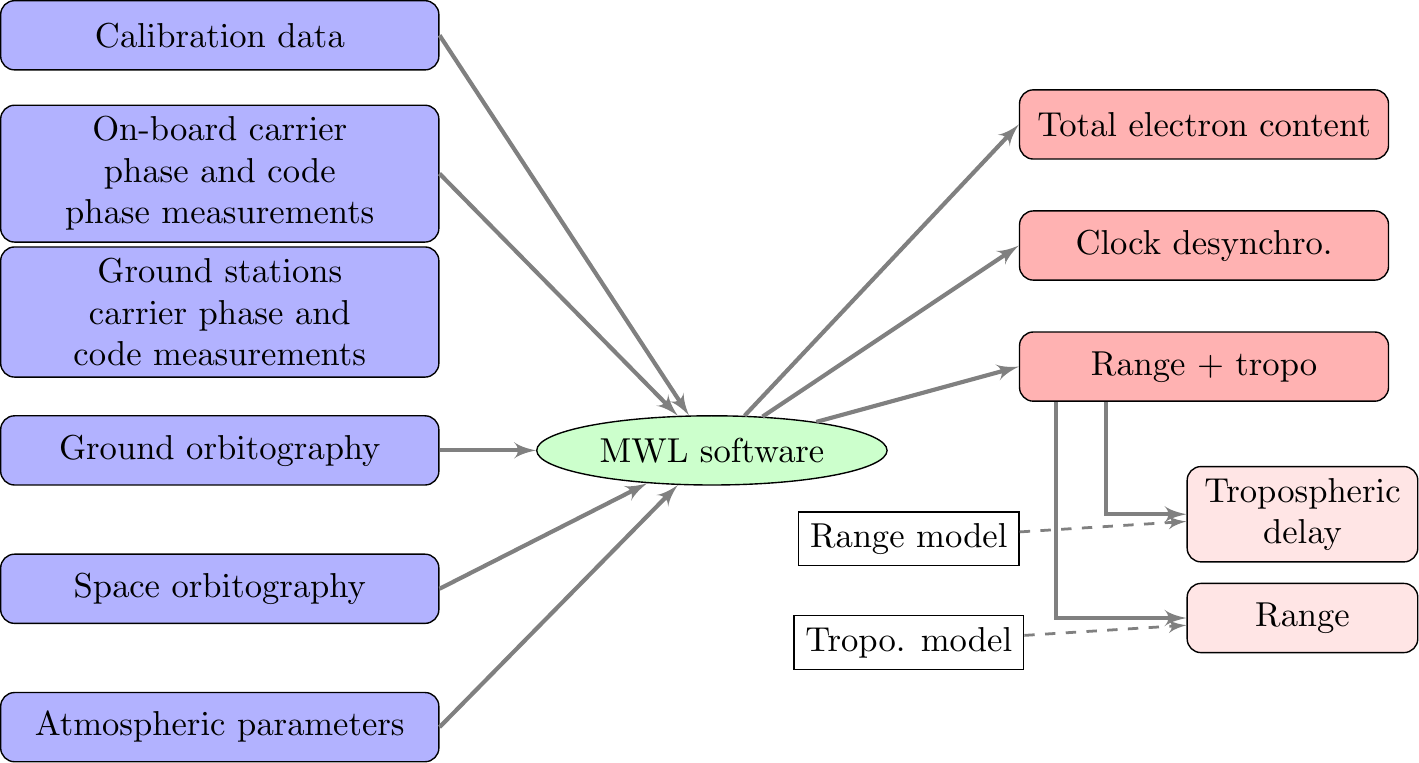} \caption{Global
    flow chart for the MWL data analysis software.}
    \label{fig:global_flowchart} \end{figure}

The goal is to transform raw data (the output of ground terminals and flight
segment) into scientific outputs, namely : the desynchronisation between a
given ground terminal clock and the flight segment, as well as some byproducts
(Total Electron Content, and a mix of range and tropospheric delays). To
achieve this, additional inputs are needed : position of the ISS and the ground
station during the pass, atmospheric parameters, and various calibration
coefficients (see figure \,\ref{fig:global_flowchart}). 

Input categories include : \begin{itemize} \item Calibration data : all
            instrumental effects have to be removed before analyzing the data.
            This is done by applying calibration coefficients and/or adding a
            calibration delay to the raw data, following procedures provided by
            the manufacturer.  \item On-board and ground station raw data : for
                a given pass, data collected by the on-board terminal and the
                corresponding ground terminal are gathered from the archive and
                fed to the software. Data is sampled at 12.5 Hz, independently
                for uplink and downlink (\textsl{i.e.} there can be up to 40 ms
                delay between an uplink data point and the closest
                corresponding downlink data point).  \item Orbitography : By
                    using a two-way time transfer, at first order the one-way
                    propagation time of the signal cancels out, but the
                    variation of this delay between forth and back paths does
                    not : this effect may shift the desynchronisation
                    measurement (up to 10 ns in the least favourable
                    configurations), so it should be mitigated with a priori
                    knowledge of the respective positions of the terminals,
                    \textsl{e.g.} orbitography of the ISS and position of the
                    ground terminal. Naming ground station position ``Ground
                    orbitography'' stresses the symmetry between both stations
                    : in a non-rotating frame, ground stations also are in
                    motion during the signal's propagation.  \item Atmospheric
                        parameters are used for first order estimation of the
                        tropospheric delay \end{itemize}

Scientific outputs include (see section \ref{sec:simu}): \begin{itemize} \item
            The clock desynchronization between ground and space, obtained from
            the difference between uplink and downlink Ku band measurements.
            This measures, for a given coordinate time $t$, the difference
            between the ground clock's proper time and the space clock's proper
            time.  \item The Total Electron Content, derived from the
                difference between Ku and S band downlink measurements. This is
                a by-product of the ionospheric delay determination, which is
                necessary to correct the desynchronisation product.  \item A
                    mix of range and tropospheric delays obtained from the sum
                    of uplink and downlink Ku band measurements. Modelling one
                    of those two effects allows to derive the remaining one.
            \end{itemize}

\subsection{Validating the software with simulated data}

Data analysis preparation has begun long before any ``real'' data was
available. Moreover, the flexibility of our simulation software allows us to
investigate cases that cannot be emulated on the hardware. Finally, the ability
to generate two datasets with only one parameter being changed, all others
being kept identical, is a very useful tool to validate the individual modules
of our codes.

The rather strict separation we maintained between data analysis and simulation
code development allowed us to highlight bugs in both codes efficiently.
Although we cannot rule out that a common misunderstanding of hardware
specifications leads to mutually cancelling bugs in simulation and data
analysis, we are confident that most of them have been solved this way.
Definite proof will, of course, only be available once actual hardware data
will be fed into the data analysis software. 


For each pass, a set of simulated data is generated, along with a corresponding
set of theoretical data. The former contains what we suppose will come out of
the Ground Segment, \textsl{i.e.} phase counter values as decommuted from
telemetry. The latter contains the corresponding theroretical values, which is
a list of what has been calculated at different steps in the simulation
software : scientific products but also intermediary values (\textsl{e.g.}
tropospheric delay, ionospheric delay, etc\ldots) are thus directly comparable
to the corresponding values as calculated by the data analysis software.

We cannot expect full recovery of those values : the fact that data is only
collected through counter values introduces an uncertainty on all of them. We
assume that a specific calculation is validated when, for one pass, the offset
and scatter of all data points within this pass are compatible with the
dispersion expected from this truncation.

For example, we know from the MWL measurement principle \cite{hejcTN2011} that
the main counter frequency runs at 100.1953125 MHz. This means that any timetag
derived from this counter is truncated at $\simeq$ 10 ns resolution. This
converts into a 20 ps resolution on code pseudo-range (taking advantage of the
fact that we measure the phase of a beatnote: 10 ns $\times f\beatc\co/f\co
\approx$ 10 ns $\times$ 195 kHz/100 MHz $\approx$ 20 ps). When plotting the
residuals (calculated - expected value) for each point of the pass, we expect a
random dispersion with a width of 20 ps. Moreover, code data should be
unambiguous, so we expect the theoretical value to be within this spread
(\textsl{i.e.} the 20 ps range of residuals contains 0). So, for the case of
code pseudo-range, our validation criteria should be: peak-to-peak residuals
$\leq$ 20 ps, residuals mean offset between -10 ps and +10 ps.  Similarly for
carrier measurements we expect a resolution of about 0.5 ps ($\approx$ 10 ns
$\times f\beatc\ca/f\ca \approx$ 10 ns $\times$ 729 kHz/14 GHz), with an
overall offset that is no better than 20 ps/sqrt(N) for N passes, but should be
within 0.5 ps between passes (see section~\ref{sec:cocaamb}). 

A less quantitative (and less absolute) criterion is the absence of notable
signature in the residuals : most of the time we expect the spread to be
uniform across the pass, and any visible pattern should be investigated as a
possible consequence of implementation problems.

\begin{figure}[ht] \centering \includegraphics[width=\linewidth]{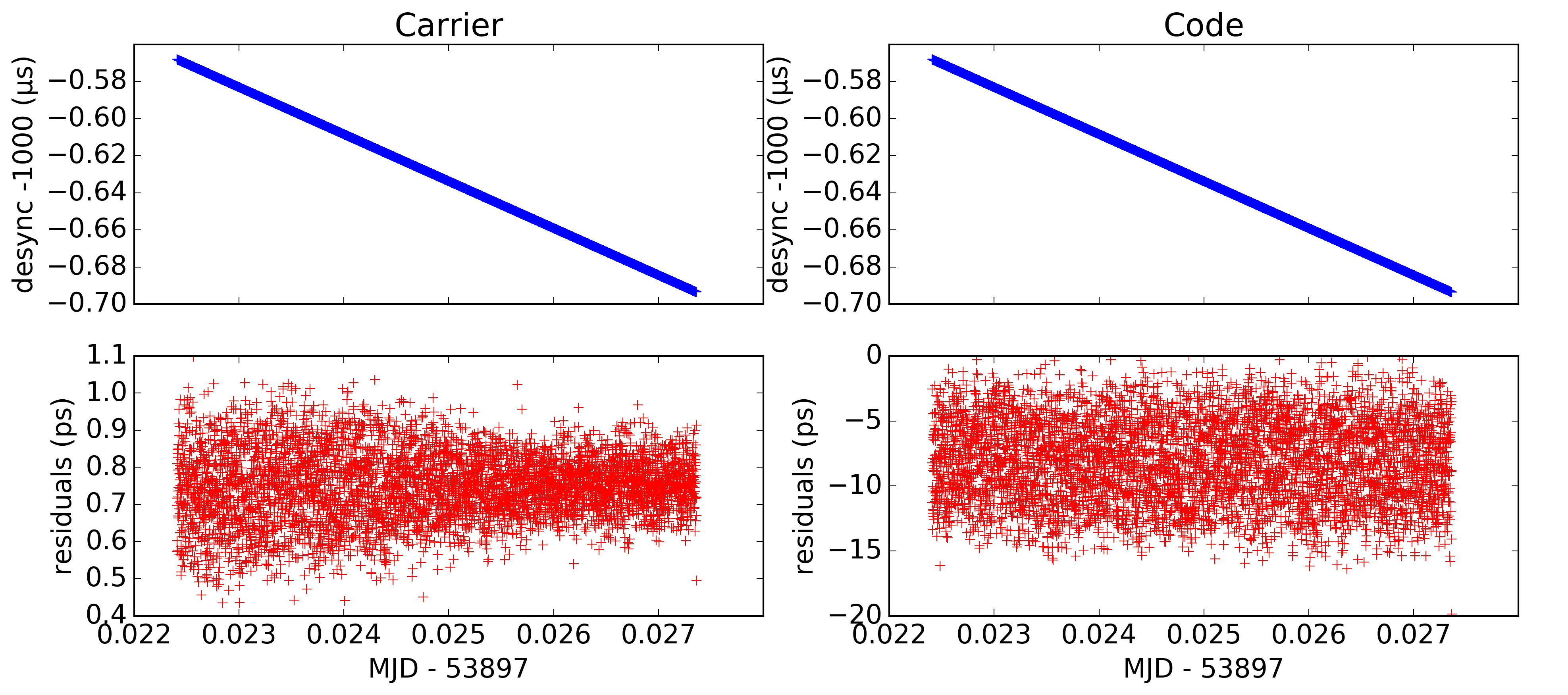}
    \caption{Typical pass data. Top : input values for desynchronisation.
    Bottom : difference between input values and processed data. Noise is
linked to the counter's resolution and represent the noise floor of the MWL.}
\label{fig:typical} \end{figure}

Figure \ref{fig:typical} shows the typical results we expect to get when
processing a single pass and compare the result to input values. Each point on
the bottom graph is a difference between the simulated (``theoretical'') value,
and the value we obtain at the end of data anlysis. Here we only show results
for the desynchronisation, but similar graphs can be drawn for any intermediate
result which is calculated for each data point (\textsl{e.g.} TEC, ionospheric
delay, etc\ldots). Code and carrier observables are obtained from distinct
datasets, with similar algorithms : as they measure the same physical
quantities we use the former to determine the latter's offset, although an
overall residual offset will remain in the carrier data (see
section~\ref{sec:cocaamb}).

It should be noted that, although the desyncronisation value (top line) spans
more than 10 ns during the pass (which is expected, due to gravity potential
and velocity difference between ground and space), we recover its value with a
peak to peak spread \textless 1 ps for carrier and \textless 20 ps for code.  


\begin{figure}[h] \centering \includegraphics[width=\linewidth]{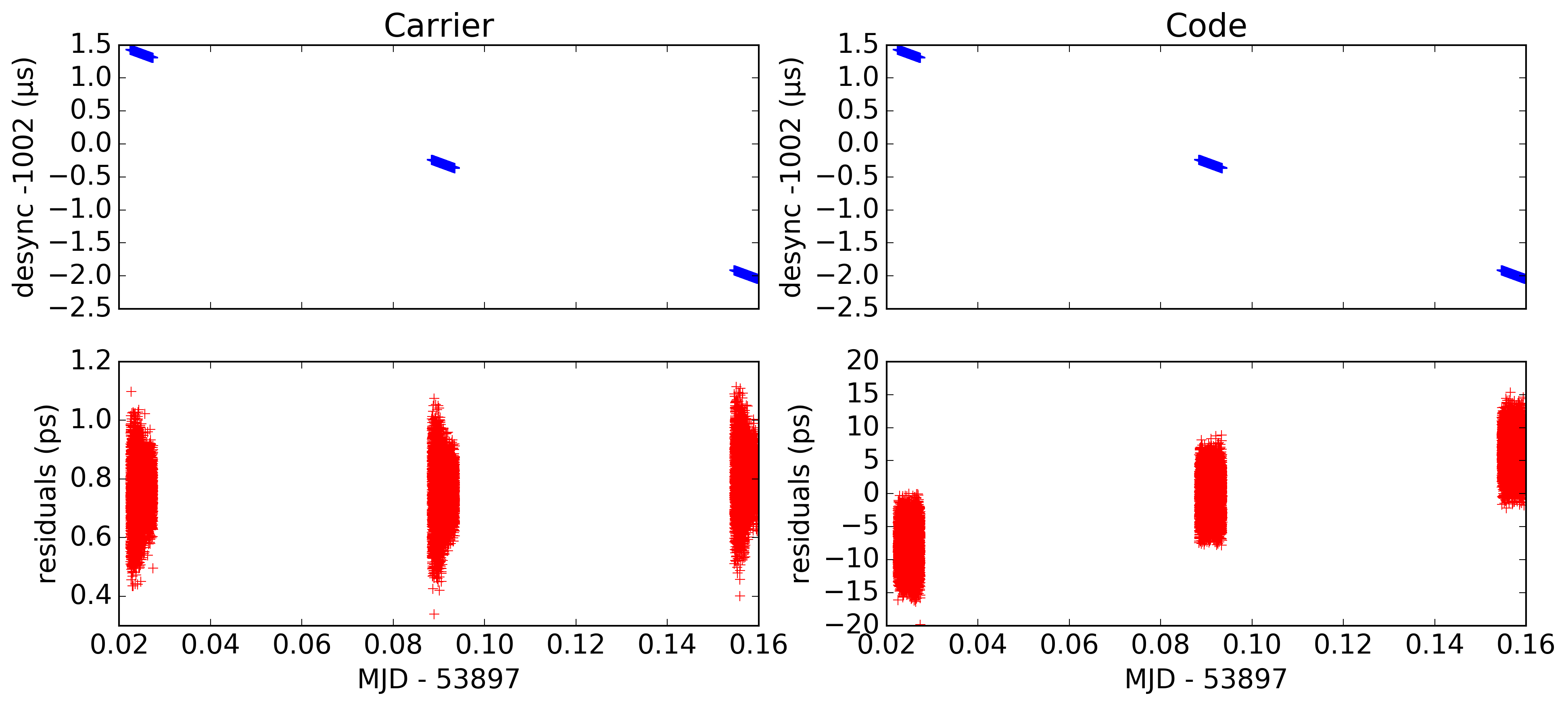}
    \caption{Similar to fig\,\ref{fig:typical} but with a larger time span,
    showing several successive ISS visiblility periods. As expected, code
    residuals show mean values in the -10/+10 ps range. Carrier residuals stay
    close to their initial value (which will be unknown in flight).  }
    \label{fig:multipass} \end{figure}

One of the key achievements needed to fulfill the ACES mission goals is the
ability to link results from one pass to the next : we will then be able to
track time differences on periods of time much longer than one pass. As code
data is unambiguous this is directly the case, albeit with a poor stability.
Carrier data is more challenging to deal with, but will bring the required
stability.

Based on our understanding of the measurement method \cite{hejcTN2011}, we have
implemented a processing algorithm that recovers the carrier phase difference,
up to an unknown offset which is set at device switch on (see
section~\ref{sec:cocaamb}). In other words, as long as both the emitter and the
receivers remain switched on, our analysis software is able to recover time
differences from the counter values with an overall offset, but a peak to peak
spread better than 1 ps at 80 ms. Figure~\ref{fig:multipass} illustrates this
capability on 3 successive passes. Here the only source of uncertainty is the
truncation due to the counters running at 100 MHz : we expect performance to be
worse in operation because of additional perturbations and noise sources.

\section{An experiment : impact of ISS orbitography errors} \label{sec5}

An immediate application of our simulation and analysis software is to check if
previous assumptions on input uncertainty impact are correct. We first had a
look at the impact of ISS orbitography uncertainty, which has already been
modeled analytically in \cite{duchayne2009}.

\subsection{Method}

The basic principle of our test is to use different orbitography data as input
for the simulation and for the data analysis. The simulation's orbitography is
then supposed to be the ``real'' one, \textsl{i.e.} the one we would get if we
had perfect knowledge of the ISS position. Then we fetch a slightly different
data as input to the analysis software : this represents the ``measured''
orbitography, \textsl{i.e.} the one that will be available when we perform the
analysis.

Modelling the uncertainty on ISS position is not straightforward, as already
noted in~\cite{duchayne2009}. Our approach here has been instead to use
existing data to estimate the position error for a given period : we took
advantage of a measurement campaign~\cite{montenbruck_orbit_2011} when both
standard Space Integrated GPS/INS (SIGI) data and more precise GNSS positioning
data (from ASN-M russian receiver) were available. Then, if we compare two
positions given for the same date, most of the error comes from the SIGI
uncertainty~\cite{Montenbruck_PrivateCom} so the difference between the two
vectors is a good estimator of the overall measurement error we would have
made, if we had only the SIGI data. 

Note that the SIGI uncertainty we assume (see figure \ref{fig:errors}) is
likely to be a pessimistic estimate of the ACES orbitography error as the ACES
payload has its own GNSS receiver which should significantly improve orbit
restitution.

Furthermore, we wanted to be able to modulate this error (\textsl{i.e.} emulate
a stronger or weaker uncertainty with respect to the real value). We therefore
wrote a tool that
\begin{itemize} \item Calculates a positional error from (SIGI - GNSS) data
        \item Projects the resulting vector in a relevant reference frame (RTN,
        see below) \item Multiplies each R, T, N component of that vector by a
            given coefficient. Below we use the same coefficient $k_{err}$ for
            all three axes.  \item Adds the resulting vector to the initial
                orbitography, providing the ``erroneous'' orbitography for the
                analysis software.  \end{itemize}

\subsection{Introduced orbitography errors}

\begin{figure}[h] \centering \includegraphics[width=\linewidth]{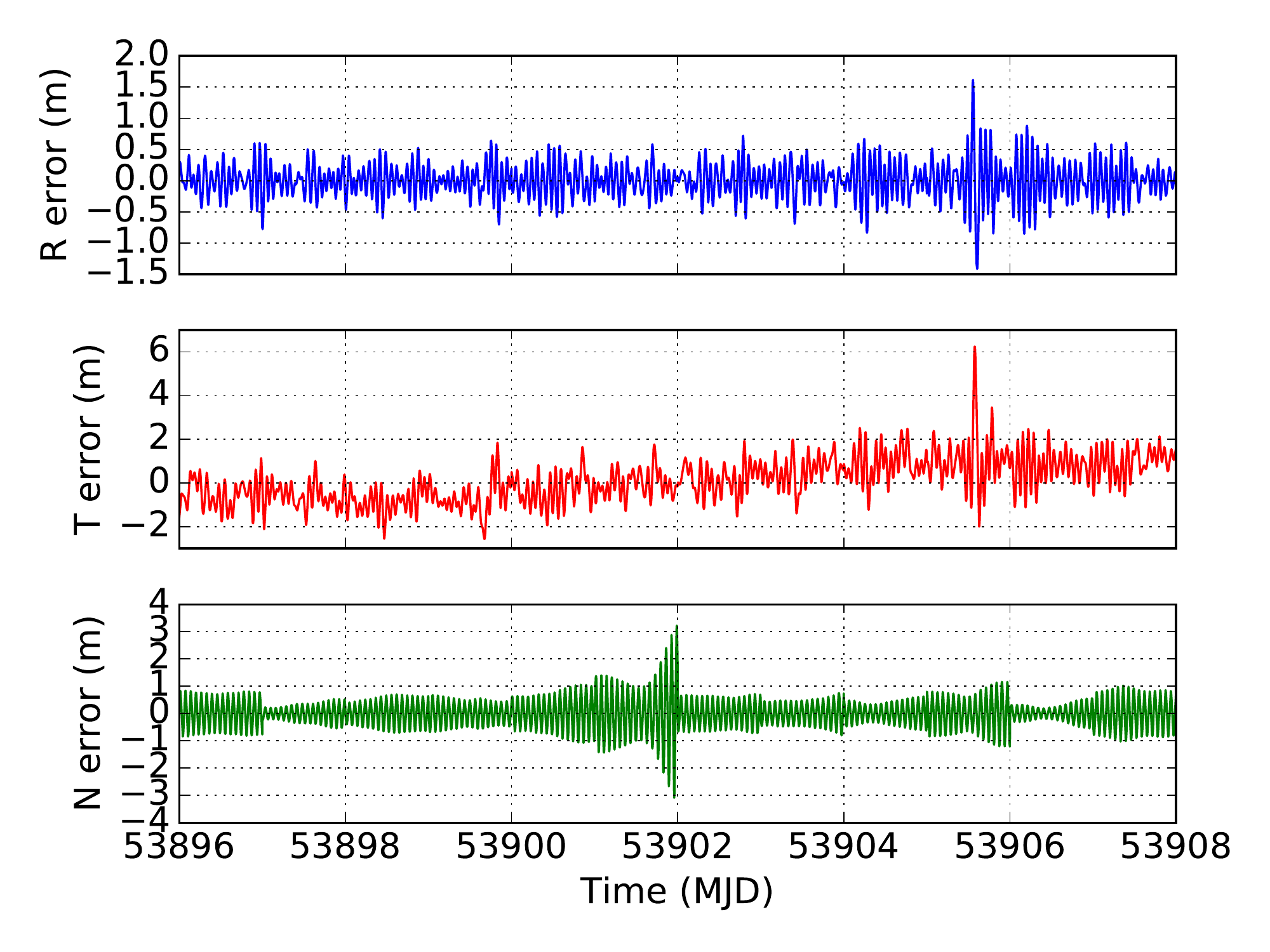}
    \caption{Difference between regular SIGI and GNSS ISS orbitography during a
    boost-free period. } \label{fig:errors} \end{figure}

We chose a period starting at MJD 53896 and ending at 53908, as altitude and
velocity data revealed no obvious event (\textsl{e.g.} altitude boost) during
this period. Resulting error is shown on figure \ref{fig:errors}.

All three orbit error components display a strong periodic component at the
orbital frequency as one might expect. The overall peak to peak error is about
8 m in T and N and somewhat less (about 3 m) in R.

\subsection{Effect on desynchronization}

\label{sec:tdevs}

\begin{figure*}[h] \centering \includegraphics[width=\linewidth]{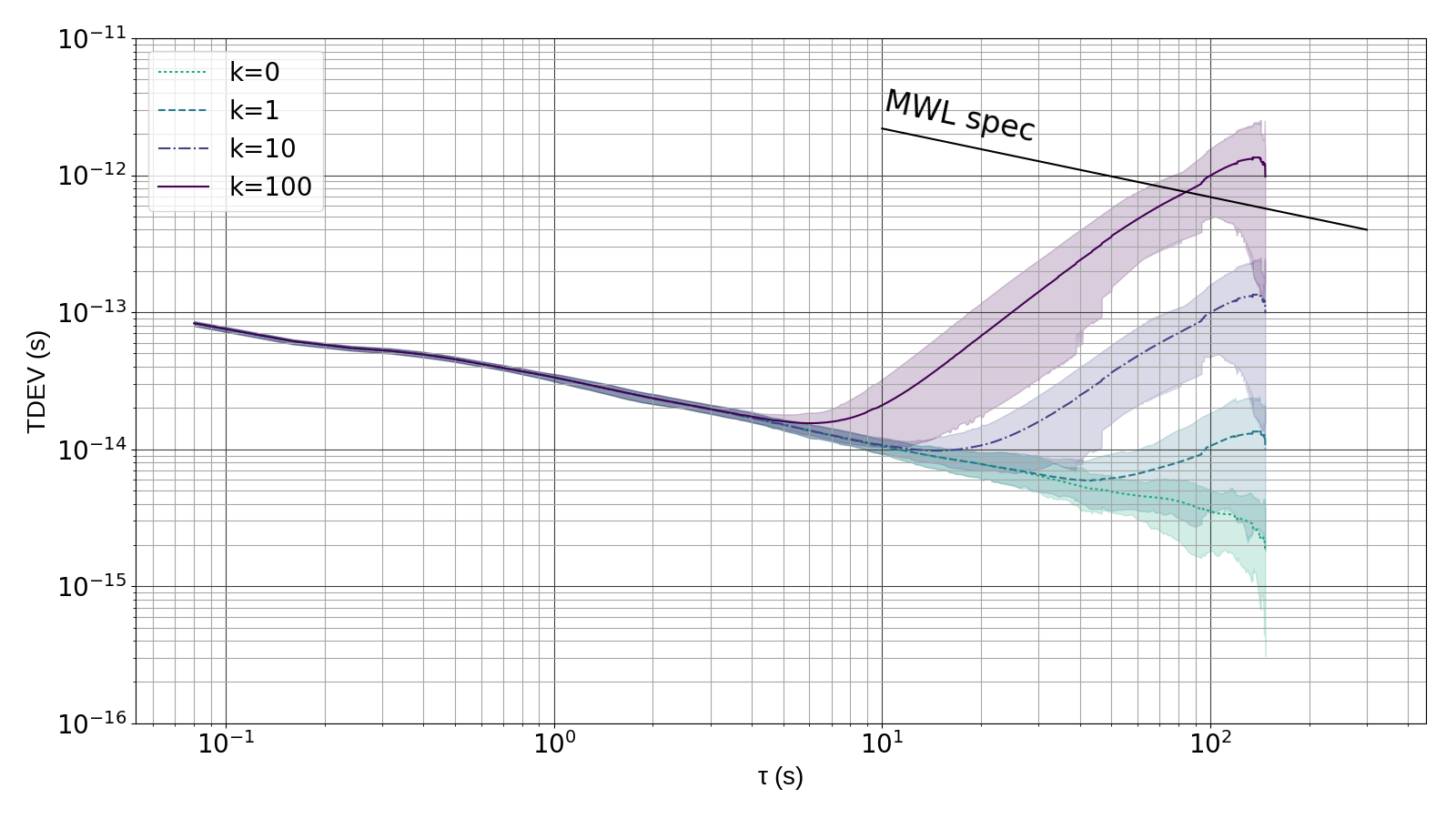}
    \caption{Effect of various uncertainty amplitude on desynchronisation
    residuals. Solid and dotted curves represent the mean value of TDEV over
all 57 passes as a function of time interval $\tau$, for each value of $k$. The
shaded enveloppes contain 80\% of the values around this mean value to
illustrate the spread. ``MWL spec'' represents the specification for the Micro
Wave Link.} \label{fig:tdevs} \end{figure*}

As a first approach we applied several levels of ISS position uncertainties,
using the same coefficient on R, T and N, on 57 consecutive passes (spanning 10
days of data).  We examined the effect of such uncertainty on the
desynchronisation residuals, and compared it to an ideal case where ISS
position is known exactly. For each coefficient ($k_{err} = 0 / 1 / 10 / 100$,
corresponding roughly to 0 m, 10 m, 100 m and 1 km peak-to-peak error), we
calculated carrier desynchronisation residuals for all 57 passes, as
illustrated on the bottom left of figure \ref{fig:typical}. Then we calculated
the corresponding TDEV for each pass. Figure \ref{fig:tdevs} represents the
Time Deviations (TDEV) calculated for each value of $k_{err}$ together with the
MWL specifications. A TDEV is calculated for each one of the 57 passes, but
only the mean value and the 80 \% interval (shaded regions) are shown.

\subsection{Discussion}

The ideal case (coefficient = 0) shows that the counter quantization, which is
the only perturbation here, introduces an error which is more than 2 orders of
magnitude better than the specification. As the coefficient grows, it can be
seen that time deviation grows accordingly towards large $\tau$ ($\simeq$ 0.5
pass length) while short term statistics is almost not affected. With an
uncertainty 100 times worse than expected, the TDEV begins to be out of
specifications. 

This leaves a comfortable margin, especially when recalling that the actual
orbitography errors should be smaller than the estimates used here because of
the additional GNSS receiver on the ACES platform. Note however, that other
perturbations will likely degrade the performance (imperfect instrumental
calibrations, uncertainty on atmospheric delays, internal delay variations,
etc\ldots).

Our results confirm the analytical estimates of \cite{duchayne2009} for the
orbitography requirements ($\approx 1$ km) coming from the specifications on
the time transfer performance. However, as shown in \cite{duchayne2009} the
requirements coming from the specifications on the integrated effect of the
gravitational redshift and second order Doppler shift are much more stringent
($\approx 10$ m). A study by numerical simulation of those effects and of their influence on the science objectives is under way and will be published in a separate paper.

\section{Conclusion}

In this article we have explained our current understanding of the necessary
steps needed to analyze the data that will be available once ACES is
operational. We also exposed some of the underlying principles and algorithms
that will allow us to perform such tasks, as well as the validation tools that
are being used for development. 

We demonstrated our ability to process typical batches of data ($\simeq$~10
days), which is an important milestone in our effort to run a Data Analysis
Center that will have to process such batches routinely during ACES operation.

As an example of application of these software tools we have used them to demonstrate that typical ISS orbitography errors ($\sim 10$~m) affect the MWL well below its performance specifications, and thus have negligible impact on the science objectives that depend on that performance.

Final software development will occur during the remaining phases prior to
launch: as soon as the definitive hardware becomes available and data schemes
freeze, we will need to fine-tune our algorithms  and adapt our flow to
possible deviations with respect to the present situation or assumptions.

\section*{Acknowledgements}

We thank Oliver Montenbruck (DLR) for providing ISS orbit data. This work was
supported by CNES.

\bibliographystyle{unsrt} \bibliography{aces_biblio_subset}

\clearpage

\appendix \section{Ionospheric delays and computation of STEC}
\label{sec:appendix} This section gives additional details on ionospheric
delays calculation and orders of magnitude. In order to deduce ionospheric
delays, we combine the two ground observables to be free of tropospheric
delays. We obtain: \begin{equation} \label{e:iono1} \begin{array}{l} \Delta
        \tau^g ( \tau^g(t^6_0) ) - \Delta \tau^g ( \tau^g (t^4_0) ) = \left[
        T_{34} - T_{56} \right]^s \\ + \left[ T_{46}^0 \right]^s - \left[
        T_{46}^0 \right]^g + \Delta^s_2 - \Delta^s_3 + \left[ \left[ \Delta^g_2
        - \Delta^g_3 \right]^t \right]^s \end{array} \end{equation}

Here we impose that $T_{46}^0 = 0$, \textsl{i.e.} both $f_2$ and $f_3$
measurements are done at the ground station at the same local time. However,
this will never be exactly zero, there will be a remaining $\delta T_{46}^0$
introducing a timing error $\delta T \simeq  \left( \epsilon_s (t_4^0) -
\epsilon_g (t_4^0) \right) \delta T_{46}^0$.  With a required accuracy $\delta
T \lesssim 0.3$~ps, we obtain the following constraint:
\begin{equation} \nonumber \delta T_{46}^0 \lesssim 0.9 \ \textrm{ms} .
\end{equation}

We expect that $|T_{34} - T_{56}| \lesssim 100$~ns
(see~\cite{duchayne:Thesis:2003}); therefore we can neglect the coordinate to
proper time transformation in equation \eqref{e:iono1}. We can also neglect
this transformation for the delays. Then equation \eqref{e:iono1} is equivalent
to:
\begin{equation} \label{e:iono3} \Dtc^g (t_6^0) - \Dtc^g (t_4^0) = T_{34} -
    T_{56} \end{equation}
From equations \eqref{eq:T34}, \eqref{eq:T56} and~\eqref{e:iono3} we obtain:
\begin{equation} \label{eq:iono2} \Delta^{\text{iono}}_{56} (f_3) -
    \Delta^{\text{iono}}_{34} (f_2) = \Dtc^g (t_4^0) - \Dtc^g (t_6^0) +
    \dfrac{R_{34} - R_{56}}{c} \end{equation} where we neglected the difference
of the Shapiro delays between the two downlinks, which can be shown to be
completely negligible.

Now we can calculate $S$ the Slant Total Electron Content~(STEC). The
ionospheric delay affects oppositely code~(co) and carrier~(ca) and may be
approximated as follows~(with S.I. units):
\begin{eqnarray} \label{e:ionod1} \Delta\cod^{\text{iono}} (f) &=
    \frac{40.308}{cf^2} S +\frac{7527}{f^3}\int N_e \left( \vec{B} \cdot
    \vec{k} \right) \textrm{d} L\\ \Delta\cad^{\text{iono}} (f) &=
    -\frac{40.308}{cf^2} S -\frac{7527}{2 f^3} \int N_e \left( \vec{B} \cdot
    \vec{k} \right) \textrm{d}L \label{e:ionod2} \end{eqnarray}
where $N_e$ is the local electron density along the path, STEC $S = \int N_e
\textrm{d}L$, $\vec{B}$ is the Earth's magnetic field and $\vec{k}$ the unit
vector along the direction of signal propagation. It has been shown that higher
order frequencies effect can be neglected for the determination of
desynchronization~\cite{duchayne:Thesis:2003}.

We suppose that for a triplet of observables \begin{equation*} \{ \Delta \tau^s
    (\tau^s(t_2^0)), \Delta \tau^g (\tau^g(t_4^0)), \Delta \tau^g
(\tau^g(t_6^0)) \} \end{equation*} the direction of signal propagation and of
the magnetic field along the line of sight do not change, and $|\vec{B}| \simeq
B_0$. Then:
\begin{eqnarray} \Delta\cod^{\text{iono}} (f) &= \frac{40.308}{cf^2} S \left( 1
    + \frac{7527 c}{40.308 f} B_0 \cos \theta_0 \right) \\
    \Delta\cad^{\text{iono}} (f) &= - \frac{40.308}{cf^2} S \left( 1 +
    \frac{7527 c}{80.616 f} B_0 \cos \theta_0 \right) , \end{eqnarray} where
$\theta_0$ is the angle between $\vec{B}$ and the direction of propagation of
signal $f_2$ and $f_3$. Then we obtain:
\begin{eqnarray} \left[ \Delta^{\text{iono}}_{56} (f_3) -
    \Delta^{\text{iono}}_{34} (f_2) \right]\cod &= \frac{40.308}{c} \left(
    \frac{1}{f_3^2} - \frac{1}{f_2^2} \right) S \nonumber\\ &\times \left[ 1 +
        \frac{7527 c}{40.308} \frac{f_2^3 - f_3^3}{f_2 f_3 \left(f_2^2 - f_3^2
        \right)} B_0 \cos \theta_0 \right]
\end{eqnarray}
\begin{eqnarray} \left[ \Delta^{\text{iono}}_{56} (f_3) -
    \Delta^{\text{iono}}_{34} (f_2) \right]\cad &= - \frac{40.308}{c} \left(
    \frac{1}{f_3^2} - \dfrac{1}{f_2^2} \right) S \nonumber\\ &\times \left[ 1 +
        \frac{7527 c}{80.616} \frac{f_2^3 - f_3^3}{f_2 f_3 \left(f_2^2 - f_3^2
        \right)} B_0 \cos \theta_0 \right] \end{eqnarray}
These equations, together with equation~\eqref{eq:iono2}, give the STEC~$S$.
The value of $S$ can then be used to correct the uplink ionospheric delay.

\end{document}